\documentclass[final, 5p, times, sort&compress]{elsarticle}

\usepackage{amssymb}
\usepackage{lipsum}
\usepackage{amsmath}

\journal{Journal of Magnetism and Magnetic Materials}

\begin{document}

\begin{frontmatter}

\title{Explaining all-optical switching in ferrimagnets with heavy rare-earth elements by varying the spin-flip scattering probability of Gd in Co$_x$Gd$_{100-x}$ alloys and Co/Gd bilayers}

\author[first]{Julian Hintermayr\corref{cor1}}
\ead{j.hintermayr@tue.nl}
\author[first]{Bert Koopmans}

\affiliation[first]{organization={Department of Applied Physics, Eindhoven University of Technology},%Department and Organization
            addressline={P.O. Box 513}, 
            postcode={5600 MB},
            city={Eindhoven},
            country={the Netherlands}}

\cortext[cor1]{Corresponding author}

\begin{abstract}
%% Text of abstract
Using the microscopic three temperature model, we simulate single-pulse all-optical switching (AOS) in alloys and bilayers consisting of Co and Gd. In particular, we investigate its dependence on the spin-flip probability of Gd $a_\mathrm{sf,Gd}$, a material parameter describing the strength of spin-phonon coupling. We do so to elucidate the mechanisms behind all-optical switching in systems where Co is coupled to heavy rare-earth elements with higher damping such as Tb. In alloys, our observations are twofold. First, an increase of $a_\mathrm{sf,Gd}$ leads to a broadening of the range of compositions for which AOS is observed. Second, the ideal Co content is decreased as $a_\mathrm{sf,Gd}$ is varied. For bilayers, our analysis indicates that switching is most efficient when $a_\mathrm{sf,Gd}$ takes on small values. Conversely, increasing the value of $a_\mathrm{sf,Gd}$ leads to a general suppression of AOS. Comparing alloys to bilayers, we find that AOS in alloys exhibits greater resilience to variations in $a_\mathrm{sf,Gd}$ than it does in bilayers.
\end{abstract}

\end{frontmatter}

%% \linenumbers

%% main text
\section{Introduction}
All-optical manipulation of magnetic domains by ultrashort laser pulses provides a swift and energy-efficient means of writing magnetic bits. Since its first discovery in a GdFeCo alloy~\cite{Stanciu:2007, Radu:2011}, significant efforts have been dedicated to both comprehending this phenomenon as well as the hunt for alternative material platforms that show this fascinating effect. 

The origin of AOS in Co-Gd-based alloys and multilayers is believed to stem from differences in the demagnetization rate of Co and Gd~\cite{Radu:2011}. The much slower magnetic relaxation time of Gd, in addition to exchange interactions between Co and Gd in the non-equilibrium state, leads to the emergence of a transient ferromagnetic (FM) state following the laser stimulus that facilitates the magnetization reversal. Atomistic simulations~\cite{Atxitia:2014, Chimata:2015, Moreno:2017, Jakobs:2021, Jakobs:2022, Jakobs:2022b, Liu:2023b}, phenomenological models~\cite{Mentink:2012, Davies:2020, Jakobs:2022}, as well as rate equations based on simple model Hamiltonians, often referred to as the microscopic three temperature model (M3TM)~\cite{Schellekens:2013, Beens:2019}, succeeded in reproducing this intriguing feature.

Recent studies reported on the observation of single-shot AOS in systems where instead of Gd, or---in addition to Gd---heavy RE elements with more than half-filled $4f$ shells and significant orbital angular momentum such as Tb~\cite{Felix:2020, Hu:2022, Peng:2023a, Hintermayr:2023b, Mishra:2023}, Dy~\cite{Hu:2022, Peng:2023a}, or Ho~\cite{Peng:2023}, were paired with Co. Those findings were somewhat unexpected, since those elements are expected to demagnetize on much faster timescales than Gd, owing to stronger spin--lattice coupling~\cite{Wietstruk:2011, Frietsch:2020}. Thus, the criterion of a strong discrepancy between the demagnetization rates of Co and the HM is not given for heavy REs as compared to Gd.

However, some findings from AOS measurements in these systems indicate a slower reversal process compared to Co-Gd, potentially driven by precessional switching. This implies that in systems such as Co/Tb multilayers, the criteria for achieving ultrafast (sub-ps) AOS are not fulfilled, yet other processes facilitate the reversal~\cite{Mishra:2023, Peng:2023a}.

Several atomistic studies modelled the replacement of Gd with Tb as an increase in element-specific damping of Tb compared to Gd~\cite{Ceballos:2021, Zhang:2022} since a theoretical study by Rebei~\textit{et~al.} found that the damping of RE elements in dilutely doped TM alloys arises from orbit--orbit coupling and should scale with the orbital moment of the RE element in such systems~\cite{Rebei:2006}.

Within the framework of the M3TM~\cite{Koopmans:2010}, it has been shown that in the simple case of an elemental ferromagnet, its demagnetization rate scales with the spin-flip probability $a_\mathrm{sf}$, denoting the chance that an electron-phonon scattering event results in the reversal of a spin. Furthermore, Koopmans~\textit{et~al.} found a connection between the Gilbert damping $\alpha$, a phenomenological factor typically describing magnetic relaxation in the GHz regime, and the ultrafast demagnetization rate of ferromagnets~\cite{Koopmans:2005}. This implies a proportionality between $a_\mathrm{sf}$ and $\alpha$ in the M3TM and Landau–Lifshitz–Gilbert descriptions. Similar findings have been made for atomistic simulations~\cite{Kazantseva:2008}, micromagnetic simulations~\cite{Djordjevic:2007}, and Landau-Lifshitz-Bloch approaches~\cite{Atxitia:2011}.

\begin{figure*}[!htbp]
    \centering
    \includegraphics[width=17.0cm]{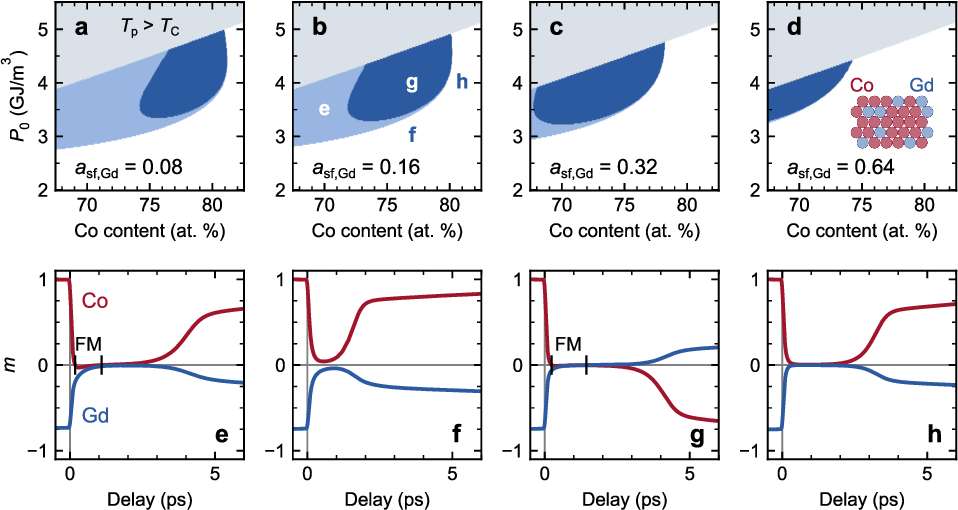}
    \caption{\textbf{a}-\textbf{d} Phase diagrams for AOS in Co$_x$Gd$_{100-x}$ alloys as a function of Co content and laser pulse energy for different values of $a_\mathrm{sf,Gd}$. White areas represent regions where no zero crossing of $m_\mathrm{Co}$ is observed. In light blue regions, $m_\mathrm{Co}$ transiently crosses zero but reverses back to its original orientation, whereas dark blue regions indicate the occurrence of AOS. The cartoon in \textbf{d} schematically illustrates the disordered alloy structure. \textbf{e}-\textbf{h} Magnetization of the Co (blue) and Gd (red) sublattice as a function of time for selected positions in the phase diagram shown in \textbf{b}. Vertical black lines in \textbf{e} and \textbf{g} indicate zero crossings of the magnetization traces and mark the emergence and subsidence of transient ferromagnetic (FM) states.}
    \label{fig:AL_PD}
\end{figure*}

Here, we mimic the effect of introducing heavy RE elements with higher magnetic damping to Co-Gd-based alloys and multilayers by increasing the spin-flip probability of Gd $a_\mathrm{sf,Gd}$ within the framework of the M3TM. Compared to atomistic simulations, this approach is computationally much less expensive, offering the possibility to easily explore large parameter spaces. We investigate the feasibility of AOS and study the ideal conditions as a function of Co:Gd balance and $a_\mathrm{sf,Gd}$. This allows us to draw comparisons not only with existing modelling work on AOS in systems with several RE elements or where magnetic damping was varied, but also helps to explain recent experimental observations.

\section{Model}

This work discusses the ultrafast magnetization dynamics in the framework of the M3TM~\cite{Koopmans:2010}. A two temperature model is used to describe interactions between the electron system, which is modelled as a spinless free electron gas, and the phonon system, which obeys the Debye model. The electron and phonon systems are assumed to be in internal equilibrium, leading to a homogeneous electron temperature $T_\mathrm{e}$ and phonon temperature $T_\mathrm{p}$. Further, the heat capacity of the spin system is neglected. At the start of the simulation, the phonon and electron temperatures are in equilibrium at ambient temperature $T_\mathrm{amb}=295~\mathrm{K}$. Ultrafast laser excitation is modelled as a transient influx of energy into the electron system, following a Gaussian profile with a fixed pulse duration of 50~fs. Additionally, a dissipation term allows energy to flow from the phonon system to the substrate with a characteristic timescale $\tau_\mathrm{D}$. The substrate is kept at $T_\mathrm{amb}$. Typical time traces of $T_\mathrm{p,e}$ and details on the system parameters are presented in the Appendix.

The magnetic Co and Gd sublattices are modelled as two antiparallel exchange-coupled spin systems with $S_\mathrm{Co}=1/2$ and $S_\mathrm{Gd}=7/2$ whose spin occupations are evaluated using the (in the case of bilayers, layered) Weiss model. At each lattice site $i$, $\mu_\mathrm{at,i}/2S_\mathrm{i}$ spins are present, where $\mu_\mathrm{at}$ (in units of the Bohr magneton $\mu_\mathrm{B}$) denotes the atomic magnetic moment and $S$ the spin quantum number.

Two processes govern the magnetization dynamics in the laser-induced non-equilibrium state: (i)~Elliot-Yafet electron-phonon scattering events have a certain probability $a_\mathbf{sf}$ of triggering a spin-flip event, resulting in transfer of angular momentum from the spin to the lattice system. The change in magnetization due to these events is implemented as described by the standard M3TM~\cite{Koopmans:2010}. (ii)~Similarly, exchange scattering between Co and Gd electrons carrying opposite spin can lead to a mutual transfer of spin angular momentum between the magnetic sublattices. We follow the same formalism as used in Ref.~\cite{Beens:2019}, where further details are provided. The used modelling parameters are listed in the Appendix.

%-----------------------------------------------------------%
\section{Results}
\subsection{Co$_x$Gd$_{100-x}$ alloys}
We start by investigating the effect of increasing $a_\mathrm{sf,Gd}$ on the AOS characteristics in amorphous Co$_x$Gd$_{100-x}$ alloys. Figure~\ref{fig:AL_PD} shows the switching diagrams for such alloys (shown schematically in the inset of Fig.~\ref{fig:AL_PD}~\textbf{d}) as a function of the atomic Co content and the laser pulse energy density $P_0$ for different values of $a_\mathrm{sf,Gd}$. The diagram shown in Fig.~\ref{fig:AL_PD}~\textbf{a} corresponds to the case where the literature value of $a_\mathrm{sf,Gd}=0.08$ is taken~\cite{Koopmans:2010}, resulting in the same diagram as obtained by Beens~\textit{et~al.} in Ref.~\cite{Beens:2019}. In Figs.~\ref{fig:AL_PD}~\textbf{a}-\textbf{d}, $a_\mathrm{sf,Gd}$ is progressively doubled up to $0.64$. The meaning of the differently colored regions is explained using the phase diagram in Fig.~\ref{fig:AL_PD}~\textbf{b} and individual magnetization traces of Co and Gd in Fig.~\ref{fig:AL_PD}~\textbf{e}-\textbf{h}. In white regions, the laser energy is not sufficient to drive the magnetization of the Co sublattice through zero, resulting in demagnetization traces similar to the one in Fig.~\ref{fig:AL_PD}~\textbf{f}. In regions colored in light blue, the Co magnetization transiently crosses zero, leading to a FM state, but relaxes back to its original orientation. Dark blue regions indicate a fully reversed state (determined by whether or not the Co magnetization has reversed its sign 100~ps after the laser stimulus), following magnetization traces similar to that in Fig.~\ref{fig:AL_PD}~\textbf{g}. Grey regions indicate scenarios where $T_\mathrm{p}$ crosses the Curie temperature $T_\mathrm{C}$ which is known to lead to a thermally demagnetized state in practice~\cite{Gorchon:2016}. In Fig.~\ref{fig:AL_PD}~\textbf{h} the result of laser pulse excitation at a high fluence in a strongly Co-dominated region is illustrated. Using a laser energy that would lead to AOS in regions with more Gd quenches both Co and Gd sublattices close to zero, yet no transient FM state is observed, due to the fact that reduced Gd content lowers the exchange scattering rate and slows down the Co demagnetization.

\begin{figure}[!htbp]
    \centering
    \includegraphics[width=8.6cm]{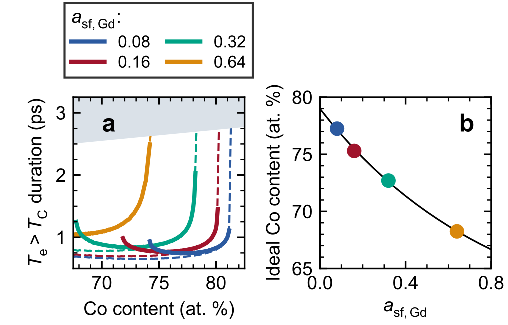}
    \caption{\textbf{a}~Duration during which $T_e$ exceeds $T_\mathrm{C}$ at the lowest fluence that results in AOS (solid lines) or a zero-crossing of the Co magnetization (dashed lines) as a function of Co content for different values of $a_\mathrm{sf,Gd}$. \textbf{b}~Co content for which AOS is observed for the largest range of $P_0$ as a function of $a_\mathrm{sf,Gd}$. The black line is a guide to the eye.}
    \label{fig:Te}
\end{figure}

\begin{figure*}[!htbp]
    \centering
    \includegraphics[width=17.0cm]{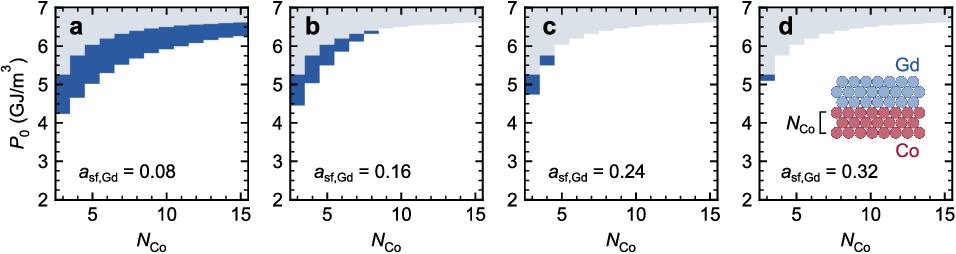}
    \caption{\textbf{a}-\textbf{d} Phase diagrams for AOS in Co/Gd bilayers as a function of number of Co layers and laser pulse energy for different values of $a_\mathrm{sf,Gd}$. The use of colors is the same as in Fig.~\ref{fig:AL_PD}. The cartoon in \textbf{d} schematically illustrates the bilayer structure.}
    \label{fig:ML_PD}
\end{figure*}

We now focus our attention on the changes in the phase diagrams as $a_\mathrm{sf,Gd}$ is increased. Two trends are apparent: First, increasing $a_\mathrm{sf,Gd}$ widens the AOS window, \textit{i.e.}, the range of compositions for which AOS is observed, towards more Gd-dominated regions. Second, the composition at which AOS occurs for the largest fluence window likewise shifts towards more Gd-dominated compositions (plotted in Fig.~\ref{fig:Te}~\textbf{b} against $a_\mathrm{sf,Gd}$). To understand this behavior, we first recall that increasing $a_\mathrm{sf,Gd}$ leads to a faster demagnetization of Gd according to the standard M3TM. As a result, it takes less time for the Gd magnetization to equilibrize with the electron system. The imbalance in spin occupation between Co and Gd is therefore also reduced, leading to lower exchange scattering rates. In Co-rich regions, where AOS was initially possible, increasing $a_\mathrm{sf,Gd}$ suppresses the reversal since the reduced exchange scattering is insufficient to drive the Co moment through zero. Conversely, more Gd-rich regions in which only a transient FM state but no reversal was observed for $a_\mathrm{sf,Gd}=0.08$ give rise to AOS for higher $a_\mathrm{sf,Gd}$. While the reason for a lack of AOS for $a_\mathrm{sf,Gd}=0.08$ in light blue regions was explained by an excess of Gd moment, the increased demagnetization rate at higher $a_\mathrm{sf,Gd}$ overcomes this limitation. The aforementioned reduced exchange scattering rate between Co and Gd caused by higher values of $a_\mathrm{sf,Gd}$ is partially counteracted by increased interactions between the two sublattices at compositions closer to the equiatomic scenario. Yet, a higher laser energy is required at the same Co content to drive the Co moment through zero for a higher $a_\mathrm{sf,Gd}$ compared to the initial case. The combination of these effects results in the shift of ideal Co content for AOS, \textit{i.e.}, the composition for which largest range of fluences leads to AOS (shown in Fig.~\ref{fig:Te}~\textbf{b}).

To gain deeper insights on the exact conditions for switching as $a_\mathrm{sf,Gd}$ is varied, we analyze the minimum duration during which $T_\mathrm{e}$ surpasses $T_\mathrm{C}$ at the threshold energies for AOS and transient FM states as a function of Co content. This duration indicates for how long the system experiences a highly non-equilibium state, in which spin transfer events occur most efficiently, before these characteristic processes can be observed. The results are extracted from the data used to generate the phase diagrams in Fig.~\ref{fig:AL_PD}~\textbf{a}-\textbf{d} and presented in Fig.~\ref{fig:Te}~\textbf{a}. Solid lines indicate the threshold behavior for AOS and broken lines the occurrence of transient FM states. For $a_\mathrm{sf,Gd}=0.08$, the duration of $T_\mathrm{e}>T_\mathrm{C}$ to induce a FM state is below 0.75~ps throughout most of the composition range. In the case of AOS, the duration is between 0.75 and 1~ps within the narrow composition pocket that allows switching. Above 80~at.~\% Co, no FM state or AOS arises, no matter the laser energy, making the duration of $T_\mathrm{e}>T_\mathrm{C}$ diverge. Note that, if $T_\mathrm{e}$ surpasses $T_\mathrm{C}$ for more than 2.5--3~ps, $T_\mathrm{p}$ will cross $T_\mathrm{C}$ as well, leading to a multidomain state, as discussed above. For higher $a_\mathrm{sf,Gd}$, the switching window in which AOS occurs not only switches towards Gd-richer regions, which also reflects in this data, but the duration where $T_\mathrm{e}>T_\mathrm{C}$ also increases monotonously. This is again caused by the fact that at higher values of $a_\mathrm{sf,Gd}$, the system needs to be driven into non-equilibrium more strongly for exchange scattering to be sufficient to drive the system into a FM state or to fully reverse it.

\subsection{Co/Gd bilayers}
We now turn our attention towards the case of bilayers. The expansion of the two-sublattice M3TM to multilayers follows the approach of Beens~\textit{et~al.} in Ref.~\cite{Beens:2019}. Here we assume that Co and Gd have 6 nearest neighboring atoms within their plane and 3 nearest neighbors in the layers above and below them. This coordination corresponds to a (0001)-oriented hexagonal closed packed (hcp) or a (111)-oriented face centered cubic (fcc) lattice structure. Since the Gd layers are only magnetized close to the Co/Gd interface, we limit the number of Gd monolayers (ML) to 3 and explore only the effect of varying the number of Co layers $N_\mathrm{Co}$. We again characterize the AOS behavior by generating switching diagrams. In contrast to the alloy case, the number of Co monolayers is varied along the x-axis. Resulting switching diagrams are presented in Fig.~\ref{fig:ML_PD} for selected values of $a_\mathrm{sf,Gd}$. Again, the first panel (Fig.~\ref{fig:ML_PD}~\textbf{a}) displays the case of the literature value for $a_\mathrm{sf,Gd}=0.08$. The choice of colors is the same as in Fig.~\ref{fig:AL_PD} with back-switching of Co being disregarded for clarity. At this point we would like to note that previous theoretical studies on bilayers have shown the magnetization reversal first occurs at the Co/Gd interfacial region before propagating through the Co layers that are more distant from the interface~\cite{Beens:2019, Beens:2019b}.

Taking a closer look at the phase diagram in Fig.~\ref{fig:ML_PD}~\textbf{a}, we find that AOS is present for all investigated $N_\mathrm{Co}$. Furthermore, the critical laser energy increases as a function of $N_\mathrm{Co}$, which can be attributed to two main factors. First, a thicker Co layer results in an elevation of $T_\mathrm{C}$ of the full stack, requiring more power to sufficiently demagnetize the system. This is also reflected in the fact that a higher laser energy is necessary to drive $T_\mathrm{p}$ above $T_\mathrm{C}$, delaying the onset of thermal multidomain formation (grey regions in phase diagrams). Second, in thicker films the Co layers further away from the Co/Gd interface need to be quenched more strongly for the switching front to fully propagate through the entire stack, further raising the critical laser energy. Upon increasing $a_\mathrm{sf,Gd}$ (Fig.~\ref{fig:ML_PD}~\textbf{b}--\textbf{d}), AOS is gradually suppressed, having the strongest influence on layers with large $N_\mathrm{Co}$. For $a_\mathrm{sf,Gd}=0.32$, AOS is only present for a small range of laser energies and the lowest number of $N_\mathrm{Co}=3$. To explain this strong dependence, we remind ourselves that switching in bilayers originates at the Co/Gd interface. Considering the coordination numbers in the model, the environment of an interfacial Co atom roughly represents that in a Co$_{78}$Gd$_{22}$ alloy. From the discussion on alloys, we recall that the ideal switching conditions at higher $a_\mathrm{sf,Gd}$ shift towards more Gd-dominated regions and AOS in Co-rich regions is suppressed. The same effect occurs at the Co/Gd interface which exhibits a fixed composition, making AOS less favorable at high $a_\mathrm{sf,Gd}$. Note that the coordination strongly depends on the degree of intermixing at the interface, which may significantly influence the switching characteristics~\cite{Beens:2019b}. Moreover, the addition of further Co layers reduces the fluence gap for which AOS is possible, leading to a full suppression at elevated values of $a_\mathrm{sf,Gd}$ and large $N_\mathrm{Co}$.

\begin{figure}[!htbp]
    \centering
    \includegraphics[width=8.6cm]{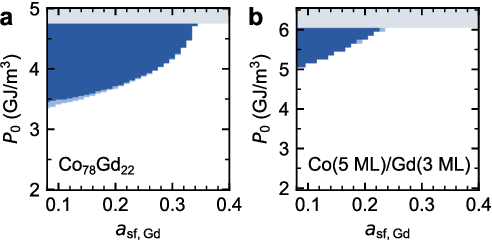}
    \caption{Switching diagrams of \textbf{a},~a Co$_{78}$Gd$_{22}$ alloy and \textbf{b},~Co(5~ML)/Gd(3~ML) as a function of $a_\mathrm{sf,Gd}$ and laser pulse energy. The use of colors is the same as in Fig.~\ref{fig:AL_PD}.}
    \label{fig:asf_PD}
\end{figure}

\section{Discussion}
Finally, we seek to bring the results of alloys and bilayers into perspective. To do so, we compute phase diagrams for representative bilayer (Fig.~\ref{fig:asf_PD}~\textbf{b}) and alloy (Fig.~\ref{fig:asf_PD}~\textbf{a}) compositions and evaluate AOS as a function of $a_\mathrm{sf,Gd}$ and $P_0$. The alloy composition was chosen in such a way that the composition resembles the local environment of Co at the Co/Gd interface in the bilayer. In both cases, switching is observed over the largest range of laser energies for the lowest values of $a_\mathrm{sf}$. Please note that this does not necessarily hold for all alloy compositions, as Figs.~\ref{fig:AL_PD} and \ref{fig:Te}~\textbf{b} illustrate that ideal switching conditions show a non-trivial dependence on Co content and $a_\mathrm{sf}$. When increasing $a_\mathrm{sf,Gd}$, AOS is gradually suppressed for both the alloy and the multilayers. However, AOS in the alloy system is more robust against changes in $a_\mathrm{sf,Gd}$ and persists over a larger range of values. This finding is in qualitative agreement with recent experiments presented in the introduction: Ceballos~\textit{et~al.} have shown that substituting Gd with Tb in amorphous Co$_{78}$Gd$_{22}$ \textit{alloys} does not suppress AOS, unless the entire Gd content is replaced by Tb~\cite{Ceballos:2021}, implying the alloy system shows a decent amount of resilience against introducing RE materials with higher damping. Atomistic modelling matched the experimental results, confirming that additional damping introduced by the heavy RE element suppresses AOS at the given composition. Further atomistic studies on AOS have similarly determined that tuning of damping parameters is crucial for the existence of AOS~\cite{Chimata:2015, Moreno:2017, Prasad:2023}. Experiments on \textit{multilayers} with Co, Tb, and Gd on the other hand showed that AOS efficiency is independent of the Tb thickness and only the balance between Co and Gd is of relevance~\cite{Hintermayr:2023b}. This is qualitatively in line with our findings that show that AOS at a pristine Co/Gd interface occurs over a wide range of fluences and layer sizes but disappears for thicker layers as $a_\mathrm{sf,Gd}$ is increased.

\section{Conclusion}
In conclusion, in an attempt to explain effects of replacing Gd by heavy RE elements with larger orbital angular momentum, we investigated the effect of increasing $a_\mathrm{sf,Gd}$ on single-shot AOS in Co$_x$Gd$_{100-x}$ alloys and Co/Gd bilayers within the framework of the M3TM. We found that the composition window in which AOS is observed in alloys shifts towards more Gd-dominated compositions as $a_\mathrm{sf,Gd}$ is raised. This shift occurs as the loss of exchange scattering suppresses AOS in Co-dominant regions but can be compensated by moving towards more Gd-dominant compositions where Co-Gd interactions are enhanced. Moreover, we showed that, as $a_\mathrm{sf,Gd}$ is raised, $T_\mathrm{e}$ spends an increasing amount of time above $T_\mathrm{C}$ before the characteristic transient FM state arises, which is likewise linked to reduced exchange scattering at higher $a_\mathrm{sf,Gd}$. For Co/Gd bilayers, switching is most efficient at low $a_\mathrm{sf,Gd}$, whereas increasing it leads to a significant suppression of AOS, especially for thicker Co layers. Comparing alloy against bilayers, AOS in alloys proved to be more robust against variations in $a_\mathrm{sf,Gd}$ than bilayers. This is caused by the fact that the interfacial region in multilayers, across which the crucial coupling between Co and Gd takes place, exhibits a fixed composition which only matches the requirements for AOS for a limited range of $a_\mathrm{sf,Gd}$. Our findings shed new light on recent experimental findings on AOS in systems with RE elements with higher magnetic damping and could help in the design of new materials for ultrafast AOS.

%--------------------------------------------------------------------%
\section*{Acknowledgment}
This project has received funding from the European Union’s Horizon 2020 research and innovation programme under the Marie Skłodowska-Curie grant agreement No 861300.

\appendix

\section{2TM and system parameters}

\begin{figure}[htbp]
    \centering
    \includegraphics{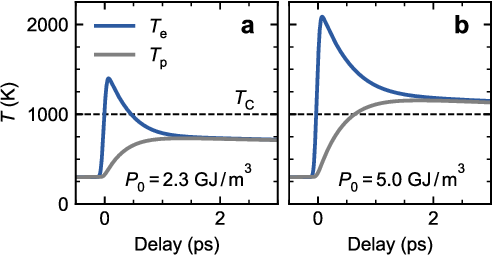}
    \caption{Solution to the system of equations given by eqs.~(\ref{eq:2TMe}) and (\ref{eq:2TMp}), describing the dynamics of $T_\mathrm{e}$ and $T_\mathrm{p}$. Two examples are given where $P_0$ suffices to \textbf{a},~transiently drive $T_\mathrm{e}$ above $T_\mathrm{C}$ and \textbf{b},~drive $T_\mathrm{p}$ above $T_\mathrm{C}$.}
    \label{fig:2TM}
\end{figure}

The time evolution of the electron temperature $T_\mathrm{e}$ and phonon temperature $T_\mathrm{p}$ is modelled according to the two-temperature model (2TM). The dynamics of the system is giving by the following system of coupled differential equations:  
\begin{align}
    \gamma T_\mathrm{e}(t) \frac{\mathrm{d}T_\mathrm{e}(t)}{\mathrm{d}t} &= g_\mathrm{ep}\left( T_\mathrm{p}(t) - T_\mathrm{e}(t)\right) + P(t), \label{eq:2TMe} \\
    C_\mathrm{p}\frac{\mathrm{d}T_\mathrm{p}(t)}{\mathrm{d}t} &= -g_\mathrm{ep}\left( T_\mathrm{p}(t) - T_\mathrm{e}(t)\right) - C_\mathrm{p}\frac{T_\mathrm{amb}-T_\mathrm{p}(t)}{\tau_\mathrm{D}}. \label{eq:2TMp}
\end{align}
$\gamma$ denotes the proportionality factor for the electron heat capacity, $C_\mathrm{p}$ the (constant) phonon heat capacity and $g_\mathrm{ep}$ the electron-phonon coupling constant. The system is excited by a transient laser pulse, described by a Gaussian function 
\begin{equation*}
    P(t)=P_0\,\mathrm{exp}\left(-t^2/\sigma^2 \right)/\left(\sigma \sqrt{\pi}\right),
\end{equation*}
with the laser pulse energy density $P_0$ and the pulse width $\sigma$. The phonon system is allowed to lowly relax towards $T_\mathrm{amb}$ over a characteristic timescale of $\tau_\mathrm{D}$. The parameters relevant for modelling the electronic and phononic system are given in Table~\ref{tab:par_ep}. Magnetic material properties relevant for the M3TM are listed in Table~\ref{tab:par_mag}. Parameters not explicitly mentioned in the main text include the element-specific demagnetization rate $R$ and the Co-Gd coupling constant $j_\mathrm{CoGd}$. The values are the same as used in Refs.~\cite{Koopmans:2010, Beens:2019}.

Two characteristic time traces of $T_\mathrm{p,e}$ are presented in Fig.~\ref{fig:2TM}. In Fig.~\ref{fig:2TM}~\textbf{a}, $T_\mathrm{e}$ transiently crosses $T_\mathrm{C}$, which is a requirement for AOS to occur, while in Fig.~\ref{fig:2TM}~\textbf{b}, $P_0$ is increased sufficiently for $T_\mathrm{p}$ to cross $T_\mathrm{C}$, leading to thermal multidomain formation.

\begin{table}[htbp]
\caption{\label{tab:par_ep}%
Parameters used modelling interactions between laser, electron, and phonon systems.}
\centering
\begin{tabular}{l@{\qquad}l@{\quad}l}
    \hline\hline
    symbol & value & {} \\
    \hline
    $g_\mathrm{ep}$ & $4.05\cdot 10^6$ & J\,K$^{-1}$\,m$^{-3}$\,ps$^{-1}$\\
    $C_\mathrm{p}$ & $4\cdot 10^6$ & J\,K$^{-1}$\,m$^{-3}$\\
    $\gamma$ & $2\cdot 10^3$ & J\,K$^{-2}$\,m$^{-3}$\\
    $\tau_\mathrm{D}$ & 20 & ps\\
    $\sigma$ & 0.05 & ps\\
    $T_\mathrm{amb}$ & 295 & K \\
    \hline\hline
\end{tabular}
\label{tab:const_ep}
\end{table}

\begin{table}[htbp]
\caption{\label{tab:par_mag}%
Parameters used modelling the magnetic subsystems.}
\centering
\begin{tabular}{l@{\qquad}l@{\qquad}l@{\quad}l}
    \hline\hline
    symbol & Co & Gd & {} \\
    \hline
    $T_\mathrm{C}$ & 1388 & 297 & K \\
    $\mu_\mathrm{at}$ & 1.72 & 7.55 & $\mu_\mathrm{B}$ \\
    $a_\mathrm{sf}$ & 0.15 & 0.08$^{1}$ & {}\\
    $R/a_\mathrm{sf}$ & 169.31 & 1.19 & ps$^{-1}$ \\
    $S$ & 1/2 & 7/2 & {}\\
    $j_\mathrm{CoGd}$ & \multicolumn{2}{c}{-1.0} & meV\\ 
    \hline\hline
    \multicolumn{4}{l}{\footnotesize ${}^1$varied throughout the paper.}
\end{tabular}
\end{table}

%\clearpage

% \bibliographystyle{elsarticle-harv} 
% \bibliography{Ref.bib}

\begin{thebibliography}{32}
\expandafter\ifx\csname natexlab\endcsname\relax\def\natexlab#1{#1}\fi
\providecommand{\url}[1]{\texttt{#1}}
\providecommand{\href}[2]{#2}
\providecommand{\path}[1]{#1}
\providecommand{\DOIprefix}{doi:}
\providecommand{\ArXivprefix}{arXiv:}
\providecommand{\URLprefix}{URL: }
\providecommand{\Pubmedprefix}{pmid:}
\providecommand{\doi}[1]{\href{http://dx.doi.org/#1}{\path{#1}}}
\providecommand{\Pubmed}[1]{\href{pmid:#1}{\path{#1}}}
\providecommand{\bibinfo}[2]{#2}
\ifx\xfnm\relax \def\xfnm[#1]{\unskip,\space#1}\fi
%Type = Article
\bibitem[{Atxitia et~al.(2014)Atxitia, Barker, Chantrell and Chubykalo-Fesenko}]{Atxitia:2014}
\bibinfo{author}{Atxitia, U.}, \bibinfo{author}{Barker, J.}, \bibinfo{author}{Chantrell, R.W.}, \bibinfo{author}{Chubykalo-Fesenko, O.}, \bibinfo{year}{2014}.
\newblock \bibinfo{title}{Controlling the polarity of the transient ferromagneticlike state in ferrimagnets}.
\newblock \bibinfo{journal}{Phys. Rev. B} \bibinfo{volume}{89}, \bibinfo{pages}{224421}.
\newblock \DOIprefix\doi{10.1103/PhysRevB.89.224421}.
%Type = Article
\bibitem[{Atxitia and Chubykalo-Fesenko(2011)}]{Atxitia:2011}
\bibinfo{author}{Atxitia, U.}, \bibinfo{author}{Chubykalo-Fesenko, O.}, \bibinfo{year}{2011}.
\newblock \bibinfo{title}{Ultrafast magnetization dynamics rates within the {Landau-Lifshitz-Bloch} model}.
\newblock \bibinfo{journal}{Phys. Rev. B} \bibinfo{volume}{84}, \bibinfo{pages}{144414}.
\newblock \DOIprefix\doi{10.1103/PhysRevB.84.144414}.
%Type = Article
\bibitem[{Avil{\'e}s-F{\'e}lix et~al.(2020)Avil{\'e}s-F{\'e}lix, Olivier, Li, Davies, {\'A}lvaro-G{\'o}mez, Rubio-Roy, Auffret, Kirilyuk, Kimel, Rasing, Buda-Prejbeanu, Sousa, Dieny and Prejbeanu}]{Felix:2020}
\bibinfo{author}{Avil{\'e}s-F{\'e}lix, L.}, \bibinfo{author}{Olivier, A.}, \bibinfo{author}{Li, G.}, \bibinfo{author}{Davies, C.S.}, \bibinfo{author}{{\'A}lvaro-G{\'o}mez, L.}, \bibinfo{author}{Rubio-Roy, M.}, \bibinfo{author}{Auffret, S.}, \bibinfo{author}{Kirilyuk, A.}, \bibinfo{author}{Kimel, A.V.}, \bibinfo{author}{Rasing, T.}, \bibinfo{author}{Buda-Prejbeanu, L.D.}, \bibinfo{author}{Sousa, R.C.}, \bibinfo{author}{Dieny, B.}, \bibinfo{author}{Prejbeanu, I.L.}, \bibinfo{year}{2020}.
\newblock \bibinfo{title}{Single-shot all-optical switching of magnetization in {Tb}/{Co} multilayer-based electrodes}.
\newblock \bibinfo{journal}{Sci. Rep.} \bibinfo{volume}{10}, \bibinfo{pages}{5211}.
\newblock \DOIprefix\doi{10.1038/s41598-020-62104-w}.
%Type = Article
\bibitem[{Beens et~al.(2019a)Beens, Lalieu, Deenen, Duine and Koopmans}]{Beens:2019}
\bibinfo{author}{Beens, M.}, \bibinfo{author}{Lalieu, M.L.M.}, \bibinfo{author}{Deenen, A.J.M.}, \bibinfo{author}{Duine, R.A.}, \bibinfo{author}{Koopmans, B.}, \bibinfo{year}{2019}a.
\newblock \bibinfo{title}{Comparing all-optical switching in synthetic-ferrimagnetic multilayers and alloys}.
\newblock \bibinfo{journal}{Phys. Rev. B} \bibinfo{volume}{100}, \bibinfo{pages}{220409}.
\newblock \DOIprefix\doi{10.1103/PhysRevB.100.220409}.
%Type = Article
\bibitem[{Beens et~al.(2019b)Beens, Lalieu, Duine and Koopmans}]{Beens:2019b}
\bibinfo{author}{Beens, M.}, \bibinfo{author}{Lalieu, M.L.M.}, \bibinfo{author}{Duine, R.A.}, \bibinfo{author}{Koopmans, B.}, \bibinfo{year}{2019}b.
\newblock \bibinfo{title}{{The role of intermixing in all-optical switching of synthetic-ferrimagnetic multilayers}}.
\newblock \bibinfo{journal}{AIP Adv.} \bibinfo{volume}{9}, \bibinfo{pages}{125133}.
\newblock \DOIprefix\doi{10.1063/1.5129892}.
%Type = Article
\bibitem[{Ceballos et~al.(2021)Ceballos, Pattabi, El-Ghazaly, Ruta, Simon, Evans, Ostler, Chantrell, Kennedy, Scott, Bokor and Hellman}]{Ceballos:2021}
\bibinfo{author}{Ceballos, A.}, \bibinfo{author}{Pattabi, A.}, \bibinfo{author}{El-Ghazaly, A.}, \bibinfo{author}{Ruta, S.}, \bibinfo{author}{Simon, C.P.}, \bibinfo{author}{Evans, R.F.L.}, \bibinfo{author}{Ostler, T.}, \bibinfo{author}{Chantrell, R.W.}, \bibinfo{author}{Kennedy, E.}, \bibinfo{author}{Scott, M.}, \bibinfo{author}{Bokor, J.}, \bibinfo{author}{Hellman, F.}, \bibinfo{year}{2021}.
\newblock \bibinfo{title}{Role of element-specific damping in ultrafast, helicity-independent, all-optical switching dynamics in amorphous ({Gd},{Tb}){Co} thin films}.
\newblock \bibinfo{journal}{Phys. Rev. B} \bibinfo{volume}{103}, \bibinfo{pages}{024438}.
\newblock \DOIprefix\doi{10.1103/PhysRevB.103.024438}.
%Type = Article
\bibitem[{Chimata et~al.(2015)Chimata, Isaeva, K\'adas, Bergman, Sanyal, Mentink, Katsnelson, Rasing, Kirilyuk, Kimel, Eriksson and Pereiro}]{Chimata:2015}
\bibinfo{author}{Chimata, R.}, \bibinfo{author}{Isaeva, L.}, \bibinfo{author}{K\'adas, K.}, \bibinfo{author}{Bergman, A.}, \bibinfo{author}{Sanyal, B.}, \bibinfo{author}{Mentink, J.H.}, \bibinfo{author}{Katsnelson, M.I.}, \bibinfo{author}{Rasing, T.}, \bibinfo{author}{Kirilyuk, A.}, \bibinfo{author}{Kimel, A.}, \bibinfo{author}{Eriksson, O.}, \bibinfo{author}{Pereiro, M.}, \bibinfo{year}{2015}.
\newblock \bibinfo{title}{All-thermal switching of amorphous {Gd-Fe} alloys: Analysis of structural properties and magnetization dynamics}.
\newblock \bibinfo{journal}{Phys. Rev. B} \bibinfo{volume}{92}, \bibinfo{pages}{094411}.
\newblock \DOIprefix\doi{10.1103/PhysRevB.92.094411}.
%Type = Article
\bibitem[{Davies et~al.(2020)Davies, Janssen, Mentink, Tsukamoto, Kimel, van~der Meer, Stupakiewicz and Kirilyuk}]{Davies:2020}
\bibinfo{author}{Davies, C.}, \bibinfo{author}{Janssen, T.}, \bibinfo{author}{Mentink, J.}, \bibinfo{author}{Tsukamoto, A.}, \bibinfo{author}{Kimel, A.}, \bibinfo{author}{van~der Meer, A.}, \bibinfo{author}{Stupakiewicz, A.}, \bibinfo{author}{Kirilyuk, A.}, \bibinfo{year}{2020}.
\newblock \bibinfo{title}{Pathways for single-shot all-optical switching of magnetization in ferrimagnets}.
\newblock \bibinfo{journal}{Phys. Rev. Appl.} \bibinfo{volume}{13}, \bibinfo{pages}{024064}.
\newblock \DOIprefix\doi{10.1103/PhysRevApplied.13.024064}.
%Type = Article
\bibitem[{Djordjevic and M\"unzenberg(2007)}]{Djordjevic:2007}
\bibinfo{author}{Djordjevic, M.}, \bibinfo{author}{M\"unzenberg, M.}, \bibinfo{year}{2007}.
\newblock \bibinfo{title}{Connecting the timescales in picosecond remagnetization experiments}.
\newblock \bibinfo{journal}{Phys. Rev. B} \bibinfo{volume}{75}, \bibinfo{pages}{012404}.
\newblock \DOIprefix\doi{10.1103/PhysRevB.75.012404}.
%Type = Article
\bibitem[{Frietsch et~al.(2020)Frietsch, Donges, Carley, Teichmann, Bowlan, Döbrich, Carva, Legut, Oppeneer, Nowak and Weinelt}]{Frietsch:2020}
\bibinfo{author}{Frietsch, B.}, \bibinfo{author}{Donges, A.}, \bibinfo{author}{Carley, R.}, \bibinfo{author}{Teichmann, M.}, \bibinfo{author}{Bowlan, J.}, \bibinfo{author}{Döbrich, K.}, \bibinfo{author}{Carva, K.}, \bibinfo{author}{Legut, D.}, \bibinfo{author}{Oppeneer, P.M.}, \bibinfo{author}{Nowak, U.}, \bibinfo{author}{Weinelt, M.}, \bibinfo{year}{2020}.
\newblock \bibinfo{title}{The role of ultrafast magnon generation in the magnetization dynamics of rare-earth metals}.
\newblock \bibinfo{journal}{Sci. Adv.} \bibinfo{volume}{6}, \bibinfo{pages}{eabb1601}.
\newblock \DOIprefix\doi{10.1126/sciadv.abb1601}.
%Type = Article
\bibitem[{Gorchon et~al.(2016)Gorchon, Wilson, Yang, Pattabi, Chen, He, Wang, Li and Bokor}]{Gorchon:2016}
\bibinfo{author}{Gorchon, J.}, \bibinfo{author}{Wilson, R.B.}, \bibinfo{author}{Yang, Y.}, \bibinfo{author}{Pattabi, A.}, \bibinfo{author}{Chen, J.Y.}, \bibinfo{author}{He, L.}, \bibinfo{author}{Wang, J.P.}, \bibinfo{author}{Li, M.}, \bibinfo{author}{Bokor, J.}, \bibinfo{year}{2016}.
\newblock \bibinfo{title}{Role of electron and phonon temperatures in the helicity-independent all-optical switching of {GdFeCo}}.
\newblock \bibinfo{journal}{Phys. Rev. B} \bibinfo{volume}{94}, \bibinfo{pages}{184406}.
\newblock \DOIprefix\doi{10.1103/PhysRevB.94.184406}.
%Type = Article
\bibitem[{Hintermayr et~al.(2023)Hintermayr, Li, Rosenkamp, van Hees, Igarashi, Mangin, Lavrijsen, Malinowski and Koopmans}]{Hintermayr:2023b}
\bibinfo{author}{Hintermayr, J.}, \bibinfo{author}{Li, P.}, \bibinfo{author}{Rosenkamp, R.}, \bibinfo{author}{van Hees, Y.L.W.}, \bibinfo{author}{Igarashi, J.}, \bibinfo{author}{Mangin, S.}, \bibinfo{author}{Lavrijsen, R.}, \bibinfo{author}{Malinowski, G.}, \bibinfo{author}{Koopmans, B.}, \bibinfo{year}{2023}.
\newblock \bibinfo{title}{{Ultrafast single-pulse all-optical switching in synthetic ferrimagnetic {Tb/Co/Gd} multilayers}}.
\newblock \bibinfo{journal}{Appl. Phys. Lett.} \bibinfo{volume}{123}, \bibinfo{pages}{072406}.
\newblock \DOIprefix\doi{10.1063/5.0161322}.
%Type = Article
\bibitem[{Hu et~al.(2022)Hu, Besbas, Smith, Teichert, Atcheson, Rode, Stamenov and Coey}]{Hu:2022}
\bibinfo{author}{Hu, Z.}, \bibinfo{author}{Besbas, J.}, \bibinfo{author}{Smith, R.}, \bibinfo{author}{Teichert, N.}, \bibinfo{author}{Atcheson, G.}, \bibinfo{author}{Rode, K.}, \bibinfo{author}{Stamenov, P.}, \bibinfo{author}{Coey, J.M.D.}, \bibinfo{year}{2022}.
\newblock \bibinfo{title}{{Single-pulse all-optical partial switching in amorphous {Dy}$_x${Co}$_{1-x}$ and {Tb}$_x${Co}$_{1-x}$ with random anisotropy}}.
\newblock \bibinfo{journal}{Appl. Phys. Lett.} \bibinfo{volume}{120}, \bibinfo{pages}{112401}.
\newblock \DOIprefix\doi{10.1063/5.0077226}.
%Type = Article
\bibitem[{Jakobs and Atxitia(2022a)}]{Jakobs:2022b}
\bibinfo{author}{Jakobs, F.}, \bibinfo{author}{Atxitia, U.}, \bibinfo{year}{2022}a.
\newblock \bibinfo{title}{Bridging atomistic spin dynamics methods and phenomenological models of single-pulse ultrafast switching in ferrimagnets}.
\newblock \bibinfo{journal}{Phys. Rev. B} \bibinfo{volume}{106}, \bibinfo{pages}{134414}.
\newblock \DOIprefix\doi{10.1103/PhysRevB.106.134414}.
%Type = Article
\bibitem[{Jakobs and Atxitia(2022b)}]{Jakobs:2022}
\bibinfo{author}{Jakobs, F.}, \bibinfo{author}{Atxitia, U.}, \bibinfo{year}{2022}b.
\newblock \bibinfo{title}{Universal criteria for single femtosecond pulse ultrafast magnetization switching in ferrimagnets}.
\newblock \bibinfo{journal}{Phys. Rev. Lett.} \bibinfo{volume}{129}, \bibinfo{pages}{037203}.
\newblock \DOIprefix\doi{10.1103/PhysRevLett.129.037203}.
%Type = Article
\bibitem[{Jakobs et~al.(2021)Jakobs, Ostler, Lambert, Yang, Salahuddin, Wilson, Gorchon, Bokor and Atxitia}]{Jakobs:2021}
\bibinfo{author}{Jakobs, F.}, \bibinfo{author}{Ostler, T.A.}, \bibinfo{author}{Lambert, C.H.}, \bibinfo{author}{Yang, Y.}, \bibinfo{author}{Salahuddin, S.}, \bibinfo{author}{Wilson, R.B.}, \bibinfo{author}{Gorchon, J.}, \bibinfo{author}{Bokor, J.}, \bibinfo{author}{Atxitia, U.}, \bibinfo{year}{2021}.
\newblock \bibinfo{title}{Unifying femtosecond and picosecond single-pulse magnetic switching in {Gd-Fe-Co}}.
\newblock \bibinfo{journal}{Phys. Rev. B} \bibinfo{volume}{103}, \bibinfo{pages}{104422}.
\newblock \DOIprefix\doi{10.1103/PhysRevB.103.104422}.
%Type = Article
\bibitem[{Kazantseva et~al.(2008)Kazantseva, Hinzke, Nowak, Chantrell, Atxitia and Chubykalo-Fesenko}]{Kazantseva:2008}
\bibinfo{author}{Kazantseva, N.}, \bibinfo{author}{Hinzke, D.}, \bibinfo{author}{Nowak, U.}, \bibinfo{author}{Chantrell, R.W.}, \bibinfo{author}{Atxitia, U.}, \bibinfo{author}{Chubykalo-Fesenko, O.}, \bibinfo{year}{2008}.
\newblock \bibinfo{title}{Towards multiscale modeling of magnetic materials: Simulations of {FePt}}.
\newblock \bibinfo{journal}{Phys. Rev. B} \bibinfo{volume}{77}, \bibinfo{pages}{184428}.
\newblock \DOIprefix\doi{10.1103/PhysRevB.77.184428}.
%Type = Article
\bibitem[{Koopmans et~al.(2010)Koopmans, Malinowski, Dalla~Longa, Steiauf, F{\"a}hnle, Roth, Cinchetti and Aeschlimann}]{Koopmans:2010}
\bibinfo{author}{Koopmans, B.}, \bibinfo{author}{Malinowski, G.}, \bibinfo{author}{Dalla~Longa, F.}, \bibinfo{author}{Steiauf, D.}, \bibinfo{author}{F{\"a}hnle, M.}, \bibinfo{author}{Roth, T.}, \bibinfo{author}{Cinchetti, M.}, \bibinfo{author}{Aeschlimann, M.}, \bibinfo{year}{2010}.
\newblock \bibinfo{title}{Explaining the paradoxical diversity of ultrafast laser-induced demagnetization}.
\newblock \bibinfo{journal}{Nat. Mater.} \bibinfo{volume}{9}, \bibinfo{pages}{259}.
\newblock \DOIprefix\doi{10.1038/nmat2593}.
%Type = Article
\bibitem[{Koopmans et~al.(2005)Koopmans, Ruigrok, Longa and de~Jonge}]{Koopmans:2005}
\bibinfo{author}{Koopmans, B.}, \bibinfo{author}{Ruigrok, J.J.M.}, \bibinfo{author}{Longa, F.D.}, \bibinfo{author}{de~Jonge, W.J.M.}, \bibinfo{year}{2005}.
\newblock \bibinfo{title}{Unifying ultrafast magnetization dynamics}.
\newblock \bibinfo{journal}{Phys. Rev. Lett.} \bibinfo{volume}{95}, \bibinfo{pages}{267207}.
\newblock \DOIprefix\doi{10.1103/PhysRevLett.95.267207}.
%Type = Article
\bibitem[{Liu et~al.(2023)Liu, Jiang, Li and Xu}]{Liu:2023b}
\bibinfo{author}{Liu, D.}, \bibinfo{author}{Jiang, C.}, \bibinfo{author}{Li, H.}, \bibinfo{author}{Xu, C.}, \bibinfo{year}{2023}.
\newblock \bibinfo{title}{{Dependence of the Gd concentration range for thermally induced magnetization switching on the intrinsic damping}}.
\newblock \bibinfo{journal}{Appl. Phys. Lett.} \bibinfo{volume}{122}, \bibinfo{pages}{202406}.
\newblock \DOIprefix\doi{10.1063/5.0135913}.
%Type = Article
\bibitem[{Mentink et~al.(2012)Mentink, Hellsvik, Afanasiev, Ivanov, Kirilyuk, Kimel, Eriksson, Katsnelson and Rasing}]{Mentink:2012}
\bibinfo{author}{Mentink, J.H.}, \bibinfo{author}{Hellsvik, J.}, \bibinfo{author}{Afanasiev, D.V.}, \bibinfo{author}{Ivanov, B.A.}, \bibinfo{author}{Kirilyuk, A.}, \bibinfo{author}{Kimel, A.V.}, \bibinfo{author}{Eriksson, O.}, \bibinfo{author}{Katsnelson, M.I.}, \bibinfo{author}{Rasing, T.}, \bibinfo{year}{2012}.
\newblock \bibinfo{title}{Ultrafast spin dynamics in multisublattice magnets}.
\newblock \bibinfo{journal}{Phys. Rev. Lett.} \bibinfo{volume}{108}, \bibinfo{pages}{057202}.
\newblock \DOIprefix\doi{10.1103/PhysRevLett.108.057202}.
%Type = Article
\bibitem[{Mishra et~al.(2023)Mishra, Blank, Davies, Avil\'es-F\'elix, Salomoni, Buda-Prejbeanu, Sousa, Prejbeanu, Koopmans, Rasing, Kimel and Kirilyuk}]{Mishra:2023}
\bibinfo{author}{Mishra, K.}, \bibinfo{author}{Blank, T.G.H.}, \bibinfo{author}{Davies, C.S.}, \bibinfo{author}{Avil\'es-F\'elix, L.}, \bibinfo{author}{Salomoni, D.}, \bibinfo{author}{Buda-Prejbeanu, L.D.}, \bibinfo{author}{Sousa, R.C.}, \bibinfo{author}{Prejbeanu, I.L.}, \bibinfo{author}{Koopmans, B.}, \bibinfo{author}{Rasing, T.}, \bibinfo{author}{Kimel, A.V.}, \bibinfo{author}{Kirilyuk, A.}, \bibinfo{year}{2023}.
\newblock \bibinfo{title}{Dynamics of all-optical single-shot switching of magnetization in {Tb/Co} multilayers}.
\newblock \bibinfo{journal}{Phys. Rev. Res.} \bibinfo{volume}{5}, \bibinfo{pages}{023163}.
\newblock \DOIprefix\doi{10.1103/PhysRevResearch.5.023163}.
%Type = Article
\bibitem[{Moreno et~al.(2017)Moreno, Ostler, Chantrell and Chubykalo-Fesenko}]{Moreno:2017}
\bibinfo{author}{Moreno, R.}, \bibinfo{author}{Ostler, T.A.}, \bibinfo{author}{Chantrell, R.W.}, \bibinfo{author}{Chubykalo-Fesenko, O.}, \bibinfo{year}{2017}.
\newblock \bibinfo{title}{Conditions for thermally induced all-optical switching in ferrimagnetic alloys: Modeling of {TbCo}}.
\newblock \bibinfo{journal}{Phys. Rev. B} \bibinfo{volume}{96}, \bibinfo{pages}{014409}.
\newblock \DOIprefix\doi{10.1103/PhysRevB.96.014409}.
%Type = Article
\bibitem[{P. and Mohanty(2023)}]{Prasad:2023}
\bibinfo{author}{P., S.P.}, \bibinfo{author}{Mohanty, J.R.}, \bibinfo{year}{2023}.
\newblock \bibinfo{title}{Single shot all-optical switching in amorphous {TbCo} and the role of element specific damping on helicity-independent all-optical switching}.
\newblock \bibinfo{journal}{J. Magn. Magn. Mater.} \bibinfo{volume}{575}, \bibinfo{pages}{170701}.
\newblock \DOIprefix\doi{https://doi.org/10.1016/j.jmmm.2023.170701}.
%Type = Article
\bibitem[{Peng et~al.(2023a)Peng, Malinowski, Gorchon, Hohlfeld, Salomoni, Buda-Prejbeanu, Sousa, Prejbeanu, Lacour, Mangin and Hehn}]{Peng:2023}
\bibinfo{author}{Peng, Y.}, \bibinfo{author}{Malinowski, G.}, \bibinfo{author}{Gorchon, J.}, \bibinfo{author}{Hohlfeld, J.}, \bibinfo{author}{Salomoni, D.}, \bibinfo{author}{Buda-Prejbeanu, L.}, \bibinfo{author}{Sousa, R.}, \bibinfo{author}{Prejbeanu, I.}, \bibinfo{author}{Lacour, D.}, \bibinfo{author}{Mangin, S.}, \bibinfo{author}{Hehn, M.}, \bibinfo{year}{2023}a.
\newblock \bibinfo{title}{Single-shot helicity-independent all-optical switching in $\mathrm{Co}/\mathrm{Ho}$ multilayers}.
\newblock \bibinfo{journal}{Phys. Rev. Appl.} \bibinfo{volume}{20}, \bibinfo{pages}{014068}.
\newblock \DOIprefix\doi{10.1103/PhysRevApplied.20.014068}.
%Type = Article
\bibitem[{Peng et~al.(2023b)Peng, Salomoni, Malinowski, Zhang, Hohlfeld, Buda-Prejbeanu, Gorchon, Verg{\`e}s, Lin, Lacour, Sousa, Prejbeanu, Mangin and Hehn}]{Peng:2023a}
\bibinfo{author}{Peng, Y.}, \bibinfo{author}{Salomoni, D.}, \bibinfo{author}{Malinowski, G.}, \bibinfo{author}{Zhang, W.}, \bibinfo{author}{Hohlfeld, J.}, \bibinfo{author}{Buda-Prejbeanu, L.D.}, \bibinfo{author}{Gorchon, J.}, \bibinfo{author}{Verg{\`e}s, M.}, \bibinfo{author}{Lin, J.X.}, \bibinfo{author}{Lacour, D.}, \bibinfo{author}{Sousa, R.C.}, \bibinfo{author}{Prejbeanu, I.L.}, \bibinfo{author}{Mangin, S.}, \bibinfo{author}{Hehn, M.}, \bibinfo{year}{2023}b.
\newblock \bibinfo{title}{In-plane reorientation induced single laser pulse magnetization reversal}.
\newblock \bibinfo{journal}{Nat. Commun.} \bibinfo{volume}{14}, \bibinfo{pages}{5000}.
\newblock \DOIprefix\doi{10.1038/s41467-023-40721-z}.
%Type = Article
\bibitem[{Radu et~al.(2011)Radu, Vahaplar, Stamm, Kachel, Pontius, D{\"u}rr, Ostler, Barker, Evans, Chantrell, Tsukamoto, Itoh, Kirilyuk, Rasing and Kimel}]{Radu:2011}
\bibinfo{author}{Radu, I.}, \bibinfo{author}{Vahaplar, K.}, \bibinfo{author}{Stamm, C.}, \bibinfo{author}{Kachel, T.}, \bibinfo{author}{Pontius, N.}, \bibinfo{author}{D{\"u}rr, H.A.}, \bibinfo{author}{Ostler, T.A.}, \bibinfo{author}{Barker, J.}, \bibinfo{author}{Evans, R.F.L.}, \bibinfo{author}{Chantrell, R.W.}, \bibinfo{author}{Tsukamoto, A.}, \bibinfo{author}{Itoh, A.}, \bibinfo{author}{Kirilyuk, A.}, \bibinfo{author}{Rasing, T.}, \bibinfo{author}{Kimel, A.V.}, \bibinfo{year}{2011}.
\newblock \bibinfo{title}{Transient ferromagnetic-like state mediating ultrafast reversal of antiferromagnetically coupled spins}.
\newblock \bibinfo{journal}{Nature} \bibinfo{volume}{472}, \bibinfo{pages}{205}.
\newblock \DOIprefix\doi{10.1038/nature09901}.
%Type = Article
\bibitem[{Rebei and Hohlfeld(2006)}]{Rebei:2006}
\bibinfo{author}{Rebei, A.}, \bibinfo{author}{Hohlfeld, J.}, \bibinfo{year}{2006}.
\newblock \bibinfo{title}{Origin of increase of damping in transition metals with rare-earth-metal impurities}.
\newblock \bibinfo{journal}{Phys. Rev. Lett.} \bibinfo{volume}{97}, \bibinfo{pages}{117601}.
\newblock \DOIprefix\doi{10.1103/PhysRevLett.97.117601}.
%Type = Article
\bibitem[{Schellekens and Koopmans(2013)}]{Schellekens:2013}
\bibinfo{author}{Schellekens, A.J.}, \bibinfo{author}{Koopmans, B.}, \bibinfo{year}{2013}.
\newblock \bibinfo{title}{Microscopic model for ultrafast magnetization dynamics of multisublattice magnets}.
\newblock \bibinfo{journal}{Phys. Rev. B} \bibinfo{volume}{87}, \bibinfo{pages}{020407}.
\newblock \DOIprefix\doi{10.1103/PhysRevB.87.020407}.
%Type = Article
\bibitem[{Stanciu et~al.(2007)Stanciu, Hansteen, Kimel, Kirilyuk, Tsukamoto, Itoh and Rasing}]{Stanciu:2007}
\bibinfo{author}{Stanciu, C.D.}, \bibinfo{author}{Hansteen, F.}, \bibinfo{author}{Kimel, A.V.}, \bibinfo{author}{Kirilyuk, A.}, \bibinfo{author}{Tsukamoto, A.}, \bibinfo{author}{Itoh, A.}, \bibinfo{author}{Rasing, T.}, \bibinfo{year}{2007}.
\newblock \bibinfo{title}{All-optical magnetic recording with circularly polarized light}.
\newblock \bibinfo{journal}{Phys. Rev. Lett.} \bibinfo{volume}{99}, \bibinfo{pages}{047601}.
\newblock \DOIprefix\doi{10.1103/PhysRevLett.99.047601}.
%Type = Article
\bibitem[{Wietstruk et~al.(2011)Wietstruk, Melnikov, Stamm, Kachel, Pontius, Sultan, Gahl, Weinelt, D{\"u}rr and Bovensiepen}]{Wietstruk:2011}
\bibinfo{author}{Wietstruk, M.}, \bibinfo{author}{Melnikov, A.}, \bibinfo{author}{Stamm, C.}, \bibinfo{author}{Kachel, T.}, \bibinfo{author}{Pontius, N.}, \bibinfo{author}{Sultan, M.}, \bibinfo{author}{Gahl, C.}, \bibinfo{author}{Weinelt, M.}, \bibinfo{author}{D{\"u}rr, H.A.}, \bibinfo{author}{Bovensiepen, U.}, \bibinfo{year}{2011}.
\newblock \bibinfo{title}{Hot-electron-driven enhancement of spin-lattice coupling in {Gd} and {Tb} $4f$ ferromagnets observed by femtosecond x-ray magnetic circular dichroism}.
\newblock \bibinfo{journal}{Phys. Rev. Lett.} \bibinfo{volume}{106}, \bibinfo{pages}{127401}.
\newblock \DOIprefix\doi{10.1103/PhysRevLett.106.127401}.
%Type = Article
\bibitem[{Zhang et~al.(2022)Zhang, Lin, Huang, Malinowski, Hehn, Xu, Mangin and Zhao}]{Zhang:2022}
\bibinfo{author}{Zhang, W.}, \bibinfo{author}{Lin, J.X.}, \bibinfo{author}{Huang, T.X.}, \bibinfo{author}{Malinowski, G.}, \bibinfo{author}{Hehn, M.}, \bibinfo{author}{Xu, Y.}, \bibinfo{author}{Mangin, S.}, \bibinfo{author}{Zhao, W.}, \bibinfo{year}{2022}.
\newblock \bibinfo{title}{Role of spin-lattice coupling in ultrafast demagnetization and all optical helicity-independent single-shot switching in {Gd}$_{1\ensuremath{-}x\ensuremath{-}y}${Tb}$_{y}${Co}$_{x}$ alloys}.
\newblock \bibinfo{journal}{Phys. Rev. B} \bibinfo{volume}{105}, \bibinfo{pages}{054410}.
\newblock \DOIprefix\doi{10.1103/PhysRevB.105.054410}.

\end{thebibliography}

\end{document}